# Development of Particle Flow Calorimetry

José Repond
*Argonne National Laboratory, Argonne, IL 60439, USA*

This talk reviews the development of imaging calorimeters for the purpose of applying Particle Flow Algorithms (PFAs) to the measurement of hadronic jets at a future lepton collider. After a short introduction, the current status of PFA developments is presented, followed by a review of the major developments in electromagnetic and hadronic calorimetry.

## 1. Introduction: Particle Flow Calorimetry

At a future lepton collider, such as the ILC or CLIC [1], the measurement of hadronic jets will play an important role in discovering or exploring physics beyond the current Standard Model of Particle Physics. Indeed, both the energy resolution and the mass resolution of multi-jet systems will play a major role. Of particular interest is the identification of the electroweak bosons through their hadronic decay on an event-by-event basis, requiring an energy resolution of the order of 3-4% for a wide range of jet energies.

A novel approach, named Particle Flow Algorithms (PFAs) is proposed to achieve this unprecedented jet energy resolution. FPAs attempt to measure each particle in a hadronic jet individually, using the component providing the best energy/momentum resolution. In this approach, charged particles are measured with a high-precision tracker, photons with the electromagnetic calorimeter and the remaining neutral hadronic particles in a jet with the combined electromagnetic and hadronic calorimeters. Table I shows the average fraction of the jet energy carried by these particle types and the expected single particle resolution obtained with the appropriate detector subsystem.

Table I. Particles in a jet, their average fraction of jet energy and their contribution to the overall resolution.

| Particle | Average fraction of jet energy | Measured with | Contribution to resolution [$\sigma^2$] |
|---|---|---|---|
| Charged | 65% | Tracker | Negligible |
| Photons | 25% | ECAL with 15%/√E | $0.07^2$ $E_{jet}$ |
| Neutral hadrons | 10% | Calorimeter with 50%/√E | $0.16^2 E_{jet}$ |
| Total | | | $0.18^2 E_{jet}$ |

Assuming typical electromagnetic and hadronic calorimeters with resolutions of 15%/√E and 50%/√E, respectively, the overall resolution is predicted to be a spectacular 18%/√E and is dominated by the measurement of the neutral hadrons. However, this estimation assumes that the energy deposits in the calorimeter be associated unambiguously to charged particles (and therefore to be ignored) or to neutral particles (to be measured with the calorimeter). Naturally, this will not always be possible and, indeed, minimizing the 'confusion term' due to misassignments of energy deposits constitutes the major challenge of this approach and imposes certain requirements on the detector design and in particular on the design of its calorimeters. A PFA detector requires an excellent tracker within a high magnetic field, a large inner radius of the calorimeter (to increase the distance between showers from different particles), the calorimeters to be placed inside the coil (to avoid energy losses in the inert material of the solenoid), calorimeters with extremely fine segmentation of the readout (to separate showers from charged and neutral particles), an electromagnetic calorimeter with a short Molière radius (to reduce the lateral spread of electromagnetic showers) and a hadronic calorimeter with a short interaction length (to be able to fit it inside the smallest possible coil).

## 2. Imaging Calorimeters

As stated above, the application of PFAs requires dense, compact calorimeters with extremely fine segmentation. In this approach the calorimeter volume is not divided into large 'towers', as for typical calorimeters of previous detectors, but is divided into small areas, read out individually layer-by-layer. Figure 1 shows the lateral sizes of the readout areas for some of the prototypes having been developed or built in recent years. The various designs feature cell sizes from 12 × 12 cm$^2$ to as small as 50 × 50 μm$^2$.

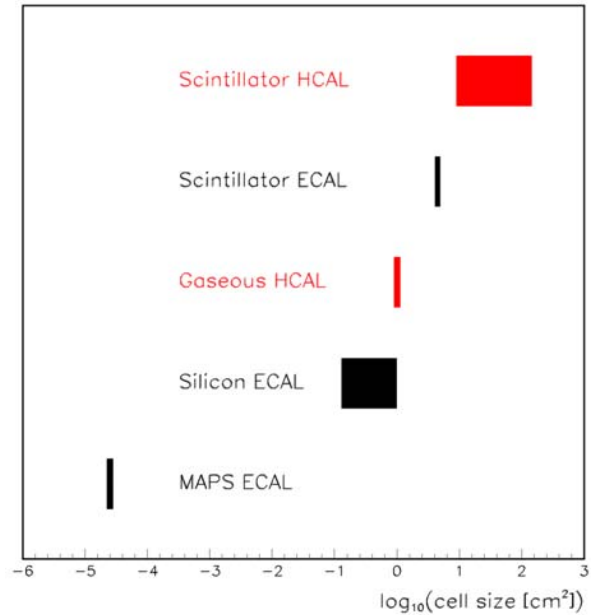

Figure 1. Cell size of some of the imaging calorimeters having being developed in recent years.

The small cell sizes naturally lead to a large number of readout channels. So, for instance the digital hadron calorimeter prototype (DHCAL) contains close to half a million readout channels, a world record in calorimetry and more than the sum of all readout channels of the four LHC calorimeters.

Apart from the ability of reconstructing every shower individually, imaging calorimeters offer several additional advantages. For instance, the direction of showers can be reconstructed with high accuracy, offering the possibility to identify particles decaying into neutrals with long decay times. The fine granularity allows identification of electromagnetic and hadronic components within individual hadronic showers and so offers the possibility to apply separate calibration constants to these sub-showers, a technique called 'software compensation'. Due to the fine longitudinal segmentation, charged particles emerging from showers can be momentum analyzed and can be measured with high precision and thus reduce the effects of leakage. Finally, energy deposits due to random noise in the calorimeter can be identified and if necessary can be eliminated.

## 3. Particle Flow Algorithms

PFAs are being developed in the context of optimizing the performance of detector concepts for the ILC and/or CLIC. Currently the best performance is obtained by the so-called PandoraPFA [2]. For simulated events in the ILD detector concept [3], a jet energy resolution of $\sigma_E/E \leq 3.8\%$ is achieved over a large range of energies. Figure 2 shows the jet energy resolution as function of the polar angle of the jet for jets of various energies. The algorithm provides an excellent resolution over most of the phase space. In the center of the detector some degradation is visible for higher energetic jets, due to energy leaking out the back of the calorimeter, while at large polar angle effects of loosing particles down the beam pipe also lead to a loss of resolution.

The development of PFAs is an open ended endeavor and is still in progress. In the current version of PandoraPFA, the resolution is in general dominated by the so-called confusion term. Also, second order corrections, such as leakage corrections or software compensation, have not yet been applied and could lead to a significant improvement of the overall performance.

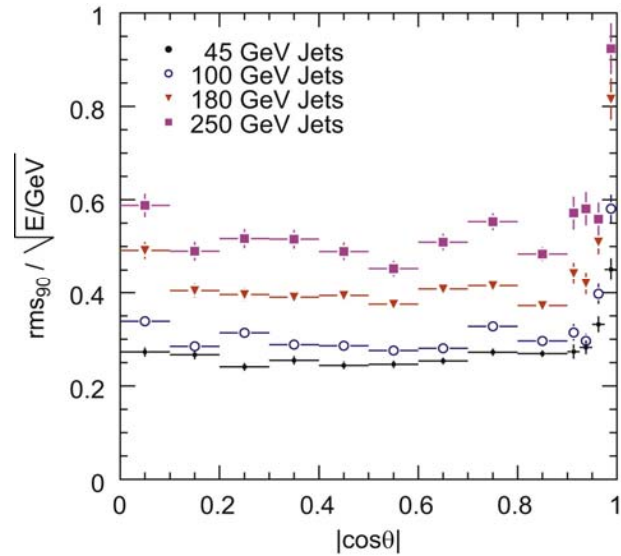

Figure 2. The jet energy resolution, defined as $\sigma_E/E = \alpha/\sqrt{E}$ plotted versus $\cos\theta_{qqbar}$ for four different values of $\sqrt{s}$. The plot shows the resolution obtained from $(Z/\gamma)^* \rightarrow qqbar$ events (q=u,d,s) generated at rest.

## 4. PFA Detector Concepts

Both the ILC and the CLIC communities are developing detector concepts optimized for the application of PFAs. In the past an alternative approach named the 4[th] concept [4] and based on a calorimeter with dual readout has been proposed, but appears to have lost momentum in recent years. The two remaining PFA detector concepts, named ILD [3] and SiD [5], are quite similar. Both feature a large solenoid, an excellent vertex tracker, a finely segmented calorimeter and a muon system on the outside of the coil. As an example, Fig. 3 shows a cross section of the SiD detector.

The major difference between the two concepts lies in the central tracker: whereas SiD features a pure Silicon tracker, ILD contains a large Time Projection Chamber in addition to several layers of Silicon trackers. These concepts were originally developed for the ILC, but have now been adapted to the requirements imposed by the higher energies of the CLIC machine.

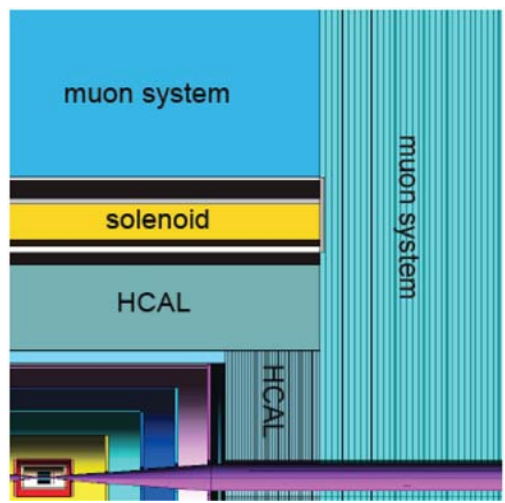

Figure 3. Cross section of the SiD detector.

## 5.  Overview of Calorimeter Developments

Imaging calorimeters for a future lepton collider are being developed in stages:

Development of the technology: in a first stage a given technology, such as MAPS, RPCs or Micromegas is being studied, developed, and adapted to calorimetry. In parallel to the sensor development an electronic readout system capable of dealing with large numbers of readout channels is being developed. This stage is often concluded with the construction and testing of a small scale prototype.

Physics prototypes: based on a deeper understanding of the chosen technology, larger scale calorimeter prototypes are being built. These physics prototypes serve as viability proof for their respective technological approach, but also provide a wealth of data on hadronic showers measured with unprecedented spatial resolution.

Technological prototypes: this next stage is expected to address all technical issues related to a given technology. These prototypes are not necessarily full scale, but are conceived as basis for the design of a first module of a colliding beam calorimeter.

Module 0: Finally, a Module 0 will aim at being the first complete module of a calorimeter of a colliding beam detector, including all the necessary bells and whistles.

Figure 4 gives an overview of the technologies currently being pursued by the ILC/CLIC community. Note that the various active media of the HCAL are being tested with both Iron and Tungsten as absorber. The boxes labeled analog/digital refer to the number of bits/channel of the readout; analog referring to several bits and digital to a single bit.

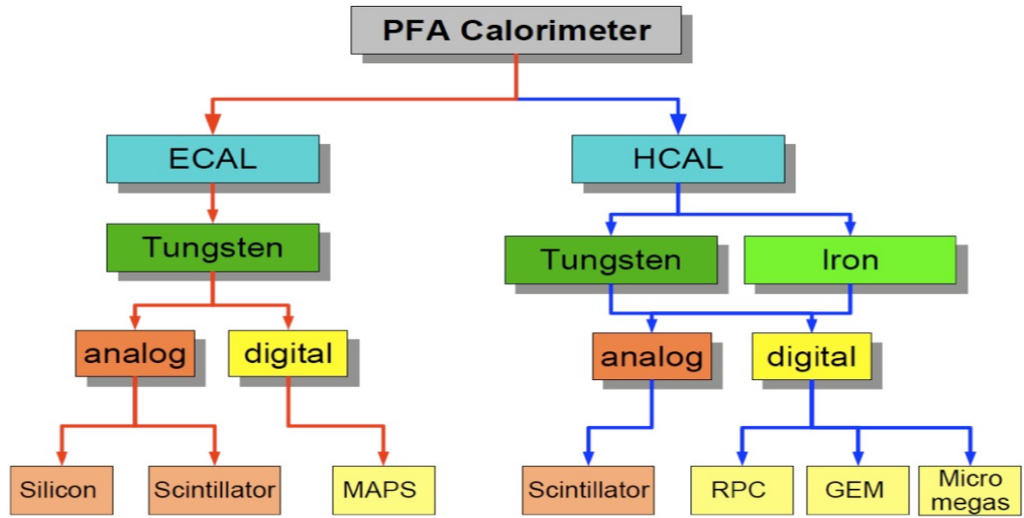

**Figure 4.** Overview of the various calorimeter technologies being pursued by the ILC/CLIC community.

Most of these projects are being carried out within the framework of the CALICE collaboration [6]. Actually, of all projects only one of the two parallel developments of a Silicon-Tungsten electromagnetic calorimeter is not associated with the CALICE collaboration. In the following we shall present the status of the most advanced of these projects.

## 6.  Silicon – Tungsten Electromagnetic Calorimeter

Silicon as active element of a calorimeter offers the possibility to be finely segmented, as required by the application of PFAs. The CALICE collaboration [6] built a first small scale prototype of a Silicon-Tungsten electromagnetic calorimeter. The device features 30 active layers interleaved by Tungsten plates with a thickness corresponding to 2/5, 4/5 and 6/5 of a radiation length each (increasing with depth into the calorimeter). The active area in each layer measures $18 \times 18$ cm$^2$ and is subdivided into $1 \times 1$ cm$^2$ pads. The calorimeter prototype counts 9,720 readout channels.

The module was extensively tested in various test beams at DESY, CERN and Fermilab, providing the first digital pictures of hadronic showers. The linearity of the response was seen to be within ±1% and the resolution was measured to be

$$\sigma_E / E = 16.6\% / \sqrt{E(GeV)} \oplus 1.1\%,$$

in excellent agreement with predictions from a complete GEANT4 [7] simulation of the setup.

The Silicon – Tungsten ECAL prototype provides an excellent tool to measure hadronic showers in great detail. As an example Fig. 5 shows the longitudinal shower shapes for 12 GeV pions. Using the fine granularity of the device the layer with the first hadronic interaction is identified and the shower shapes are measured starting with this layer, thus ignoring the preceding MIP (minimum ionizing particle) stub. The data are compared to a plethora of hadronic shower models, as implemented into the GEANT4 simulation. Also indicated are the contributions from the various particles depositing energy in a hadronic shower: protons, electrons, positrons, mesons, and others (including nuclear fragments).

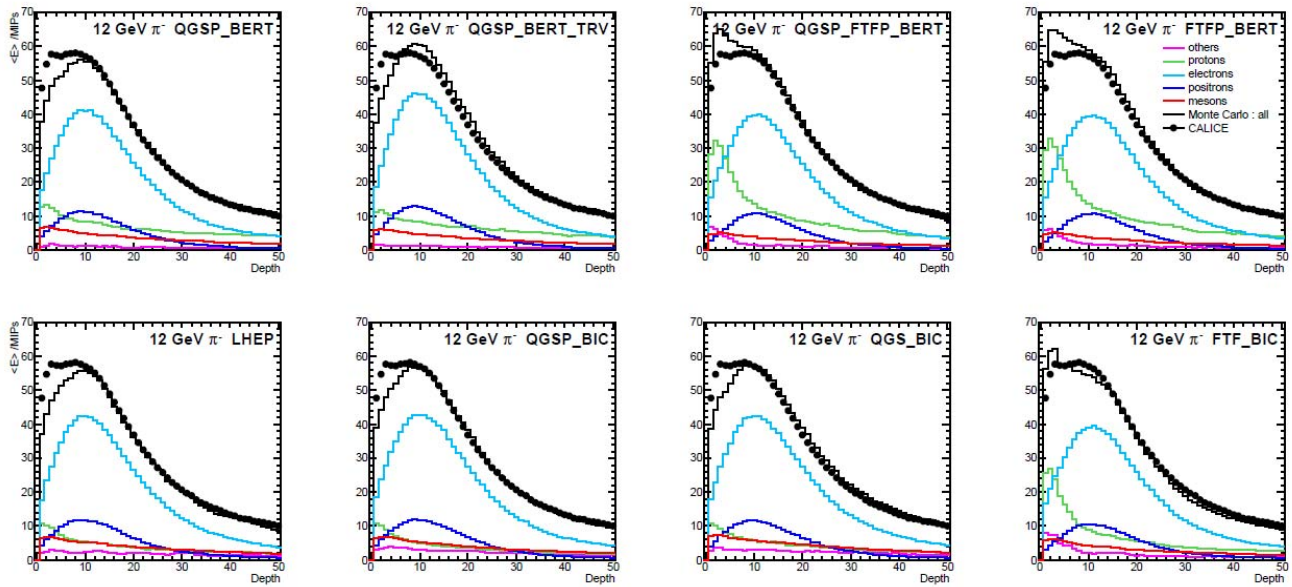

Figure 5. Longitudinal shower profiles measured in the CALICE Silicon-Tungsten ECAL prototype from the start of the hadronic shower. The black dots represent the measurements (identical in all 8 plots). The histograms are the results of GEANT4 simulations, based on various hadronic shower models. The sum of the simulation is shown as a black histogram and the colored histograms indicate the different sub-components of showers.

The power of imaging calorimetry is well illustrated by Fig.5. Measurements like these are essential to fine tune the modeling of the various components of hadronic showers. It is interesting to note, that none of the models is able to reproduce the shape close to the interaction point, with FTF_BIC perhaps best reproducing the data.

Based on the successful tests of the Silicon – Tungsten physics prototype, the CALICE collaboration initiated the design of a technical prototype. There are several significant design changes between the two prototypes: the technical prototype features smaller cell sizes, $0.5 \times 0.5$ cm$^2$, and the digitization is performed with the SPIROC chip and is part of the active layer. The thickness of the active medium is $3 - 4$ mm per layer. Figure 6 shows a sketch of the cross section through two layers of the stack. This technical prototype is currently being assembled with testing in a particle beam planned for 2012/13.

A complementary effort is staged by the SiD [5] Silicon-Tungsten group. Based on their experience with the design and construction of luminosity monitors at LEP and SLD, the group opted from the start to design and build a technical prototype. Adding to their challenge, their design pushes the current technological limits in many respects: the cell size is further reduced to hexagons with an area of 0.13 cm$^2$; the sensors are large wafers with a 6 inch diameter; and the front-end readout is based on the KPiX chip with as many as 1024 readout channels/chip. Figure 7 shows their conceptual design of a small scale prototype. The thickness of the active layer is minimized to be close to 1 mm, thus virtually maintaining the short Molière radius of Tungsten.

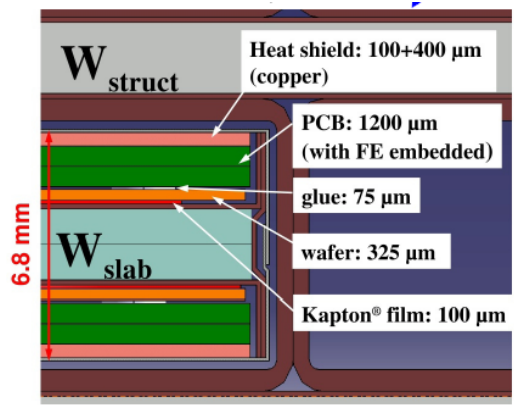

Figure 6. Cross section through two layers of the technical prototype of the CALICE Silicon-Tungsten ECAL.

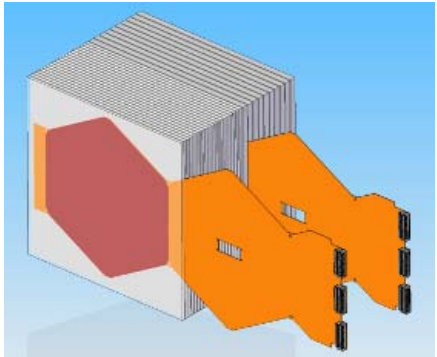

Figure 7. Conceptual design of the SiD Silicon – Tungsten electromagnetic calorimeter.

A small size calorimeter with 40 layers and 1 wafer per layer is currently being assembled. Testing in a particle beam is expected to initiate in 2013.

## 7. Scintillator – Steel Hadronic Calorimeter

The development of a Scintillator-based imaging calorimeter faced the formidable challenge of photo-detection in a strong magnetic field (thus excluding the use of traditional photomultipliers) and the readout of large number of cells within tight spatial constraints. Both challenges were met with the advent of Silicon-photomultipliers (SiPMs), i.e. multi-pixel Geiger-mode-operated avalanche photodiodes [8].

The CALICE collaboration built the first large scale calorimeter prototype using Si-PMs [9], the so-called Analog Hadron Calorimeter (AHCAL). The Scintillator tiles varied in size between $3 \times 3$ cm$^2$ in the center of a given layer up to $12 \times 12$ cm$^2$ on the outer edge. Figure 8 shows a photograph of a $1 \times 1$ m$^2$ layer with the top cover removed. The prototype consisted of 38 such layers and counted a total of about 8,000 readout channels. For this first round of prototypes the digitization of the signal was performed off the detector with a VME based system. Since 2006 the calorimeter experienced several test beams at DESY, CERN and FNAL. For these tests the active layers were interleaved with 17 mm thick steel absorber plates. Currently the prototype is undergoing its last round of testing in the CERN test beam, where the steel absorber plates have been replaced by Tungsten plates.

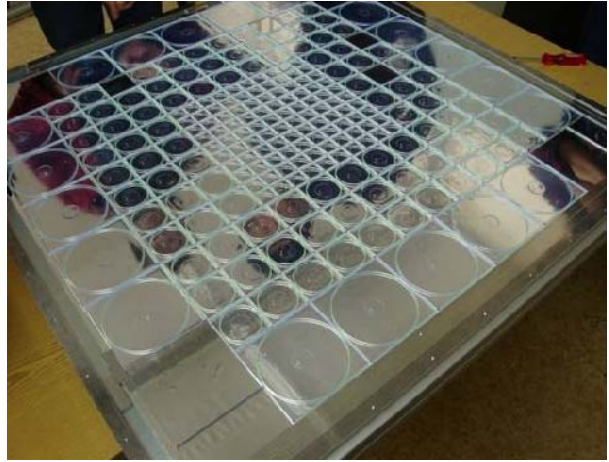

Figure 8. Photograph of an active layer of the CALICE Analog Hadron Calorimeter with the top cover removed.

A number of interesting results have emerged from the AHCAL, including a detailed measurement of the response to electrons and pions. The response to electrons and pions in this calorimeter is not equal (i.e. the calorimeter is not-compensating), leading to a measured resolution for pions of

$$\frac{\sigma_E}{E} = \frac{61.3\%}{\sqrt{E}} \oplus 2.54\%$$

Exploiting the imaging capabilities of the calorimeter, weights proportional to the energy density have been introduced to compensate for the lower response to hadrons compared to the electromagnetic component (e/h < 1). This technique is generally named software compensation. Based on this approach the resolution was significantly improved to

$$\frac{\sigma_E}{E} = \frac{49.2\%}{\sqrt{E}} \oplus 2.34\%,$$

as seen in Fig. 9. With software compensation the linearity of the response was also improved, reducing the deviations from linearity from up to 8% to below 4%. This is a prime example demonstrating the extraordinary power of imaging calorimeters.

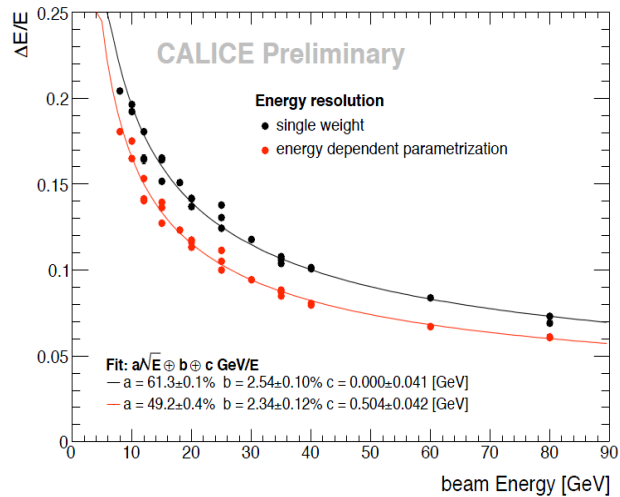

Figure 9. Hadronic resolution as measured with the CALICE Analog Hadron Calorimeter as function of pion energy. The black points and lines are before and the red points and lines are after application of software compensation. The lines are the results of fits to the stochastic plus constant terms.

In general the setup in the beam included the CALICE prototype electromagnetic calorimeter in front of the AHCAL and a Tail-Catcher-Muon-Tracker (TCMT) behind. The latter was equipped with Scintillator strips, alternating in horizontal and vertical direction and was also read out with SiPMs. Figure 10 shows the response for showers initiated by 80 GeV pions and which were selected for having started in the early part of the AHCAL. Adding up the measurements in the three parts of the setup results in a resolution of 5.4%, where the width is quoted from 90% of the events which result in the smallest root-mean-square. This measure of the width is commonly denoted as RMS90. Including only the measurements in the ECAL and the AHCAL and excluding the energy measured in the TCMT degrades the resolution to 13.1% (middle plot in Fig.10). However, applying corrections based on the identified interaction layer and/or the fraction of energy measured in the last four layers of the AHCAL is able to restore part of the degraded resolution, resulting in a width of 10.1% (right plot in Fig.10). Again, this shows the extraordinary power of imaging calorimetry.

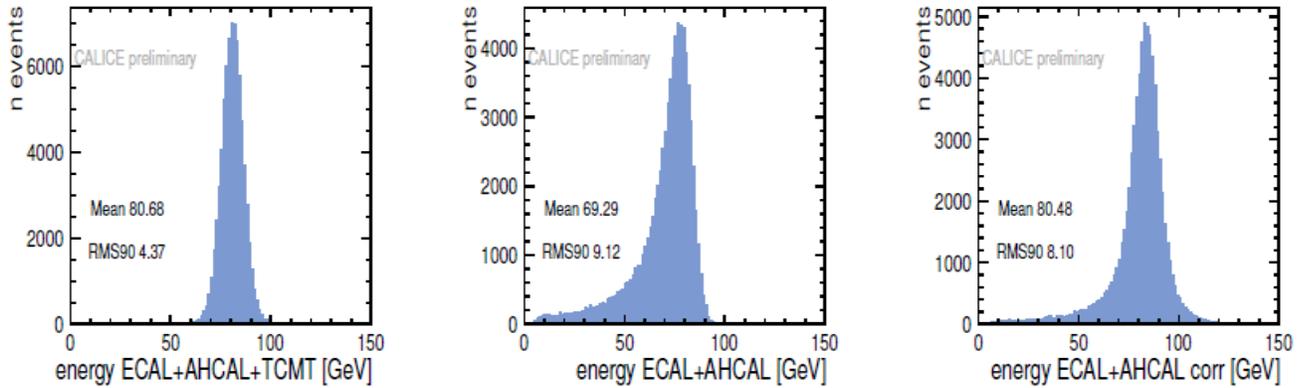

Figure 10. Response to 80 GeV pions: ECAL+AHCAL+TCMT (left), ECAL+AHCAL (middle) and ECAL+AHCAL after leakage corrections (right).

## 8. Digital Hadron Calorimeter

The concept of a Digital Hadron Calorimeter (DHCAL) is truly revolutionary. The idea is to reduce the granularity such, that a simple count of the cells with signals above a given threshold can provide a measurement of the energy of the incident particle. Thus, a low number of towers read out with high resolution in a traditional calorimeter is replaced by a large number of cells read out with a simple threshold. The energy of a particle is reconstructed to first order as being proportional to the number of cells above threshold

$$E_{rec} \propto N_{hit} \, .$$

The CALICE collaboration is pursuing several options for the active media of such a calorimeter: Resistive Plate Chambers (RPCs), Gas Electron Multipliers (GEMs), and Micromegas. Typically these devices are read out with $1 \times 1$ cm$^2$ cells and layer-by-layer.

The collaboration assembled a large RPC-based prototype with 38 layers, each measuring $1 \times 1$ m$^2$. Each layer contained three RPCs with the dimensions of $32 \times 96$ cm$^2$. Figure 11 shows a photograph of the stack, which reused the absorber structure of the AHCAL. In addition the TCMT was also equipped with RPCs (of the same design as the main stack).

The fine segmentation of the readout of the DHCAL requires the front-end electronics to be embedded into the calorimeter. Figure 12 shows a photograph of a single layer with the front cover removed. The digitization is performed by the DCAL III chip [10], developed by Fermilab and Argonne specifically for this project. Each chip reads out 64 pads and each layer contains 144 DCAL chips. The number of readout channels is 350,208 for the main stack plus 129,024 for the TCMT, for a total of 479,232 channels. This large number of readout channels constitutes a world record in calorimetry and surpasses the channel count of the calorimeters of all four LHC experiments combined.

The DHCAL was exposed to the Fermilab testbeam several times, starting in October 2010. In total more than 20 Million events were collected, such as the one shown in Fig. 13.

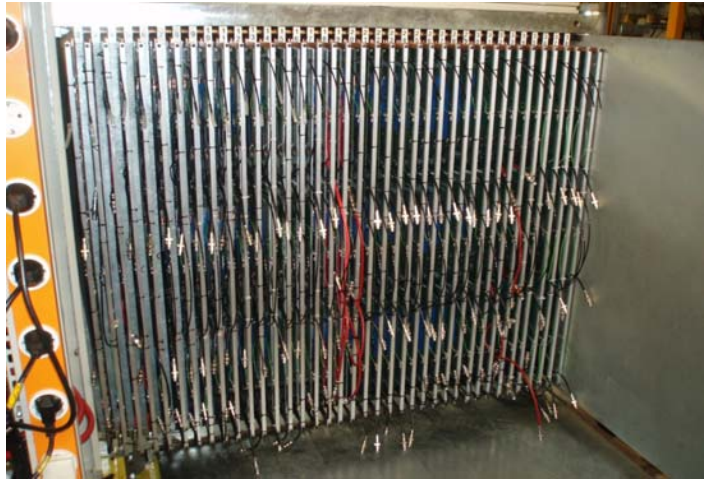

Figure 11. Photograph of the DHCAL main stack (shown before cabling) of the CALICE collaboration.

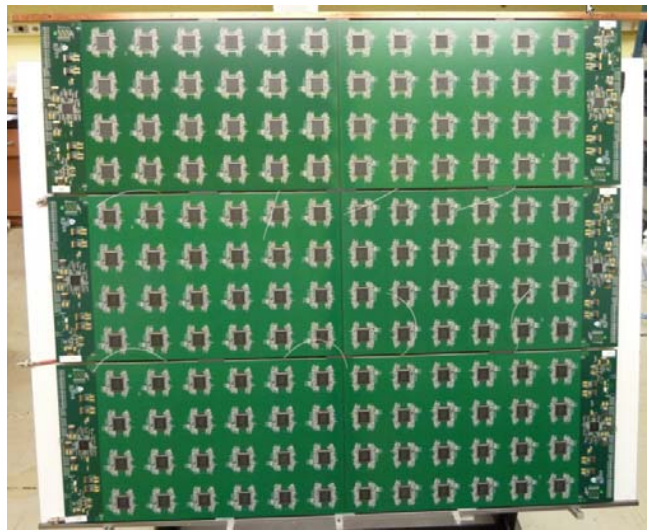

Figure 12. Photograph of a DHCAL layer with the top cover removed, showing the readout boards. The darker areas of the boards on both sides serve as data concentrators and link to a VME-based back-end readout system.

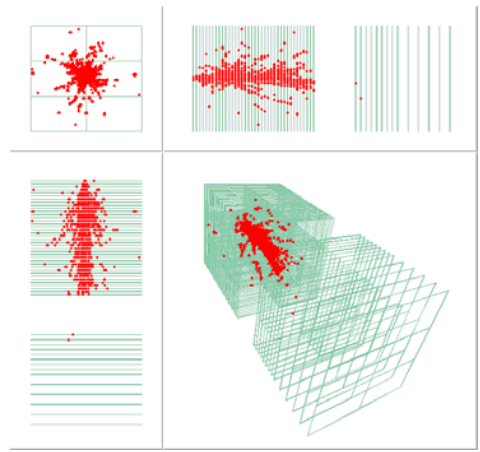

Figure 13. Event display of a shower in the DHCAL initiated by a 120 GeV proton provided by the Fermilab testbeam.

First results from the DHCAL testbeam are emerging and attest to a performance compatible with expectations [11]. Figure 14 shows the hadron resolution as function of beam energy. These results are still preliminary, having been obtained under certain assumptions, such as for instance a uniform layer-to-layer calibration. The figure shows the resolution both without and with a containment cut based on the last two layers of the main stack. In simulations based on the performance of a small prototype calorimeter a resolution of 58%/√E with a vanishing constant term was predicted. The values measured with the containment requirement are compatible with this prediction. The large constant term is in part due to variations in the layer-to-layer calibration which have not yet been dealt with.

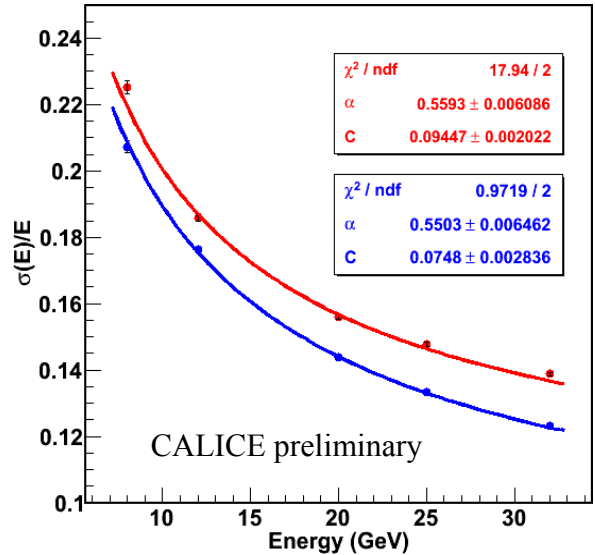

Figure 14. Hadronic resolution as measured with the DHCAL in the Fermilab testbeam. The red (blue) points and curves correspond to a data selection without (with) containment cut obtained with the last two layers of the stack.

The DHCAL was also exposed to positrons. Due to the finite size of the readout pads the response to positrons is highly non-linear and saturating at high energies. This is not considered a problem for a hadron calorimeter, as photons (or electrons) originating from the interaction point are expected to be absorbed and measured by the electromagnetic calorimeter placed in front. As a result of the non-linearity of the electromagnetic response, the ratio of the response to positrons and pions, e/h, shows interesting features, as seen in Fig. 15.

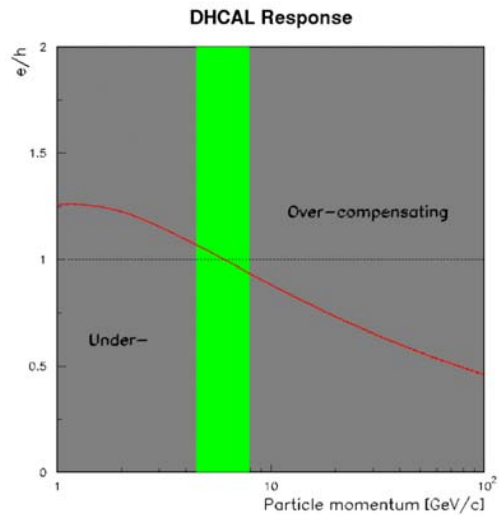

Figure 15. Ratio of electromagnetic and hadronic response as measured with the DHCAL in the Fermilab testbeam.

At low energies the DHCAL is under-compensating (e/h > 1), as is typical for most non-compensating calorimeters. As the energy increases, the finite size of the readout pads effectively suppresses the electromagnetic response achieving compensation (e/h ~ 1) for particles in the 6 – 8 GeV/c momentum range and, finally, over-compensation (e/h < 1) for higher energies. Eventually, both the effects of under-compensation and over-compensation will be diminished through the application of density-based software correction (software compensation).

## 9. The Semi-Digital RPC-based Hadron Calorimeter

The Semi-digital hadron calorimeter (SDHCAL) effort is based in Europe and is similar to the DHCAL. The active media are RPCs with an area of $1 \times 1$ m$^2$. The front-end electronics is also embedded into the calorimeter and is based on the HARDROC2b chip (also with 64 readout channels). As an extra bonus the chips are able to apply three thresholds per channel, corresponding to two bits. The front-end chip can be power pulsed to reduce power consumption in a typical linear collider environment.

A large prototype with approximately 40 layers has been assembled, see Fig. 16, and is awaiting testing in the CERN test beams.

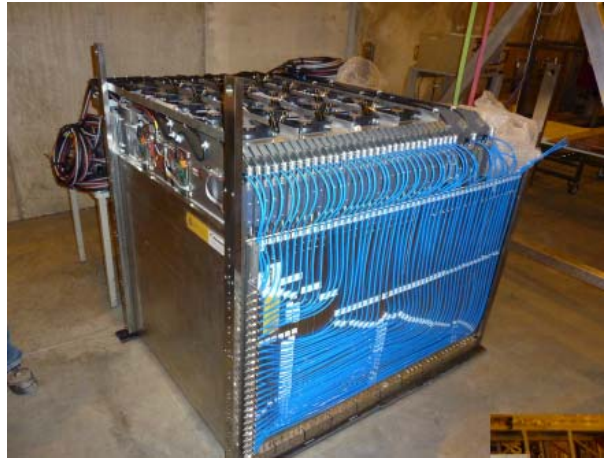

Figure 16. Photograph of the CALICE Semi-digital RPC-based hadron calorimeter (SDHCAL) prototype.

## 10. Summary

Particle Flow Algorithms are powerful tools to improve the measurement of hadronic jets at colliders. They utilize the information from both the tracking detectors and the calorimeters to measure each particle in a jet individually. PFAs have been applied at past and are being applied at present colliders to improve the measurement of jets as provided by the calorimeter alone. However, the past/present detectors have not been optimized for the application of PFAs and so the benefits of their application are somewhat limited. On the other hand, there is now a large effort underway to optimize the detectors of a future lepton collider for the application of PFAs.

In simulation studies PFAs have been shown to provide the required hadronic jet energy resolution for measurements at a future lepton collider. Based on the ILD detector concept [3] a resolution of 3-4% was obtained in the jet energy range from 45 to 250 GeV [2].

Application of PFAs requires calorimeters with unprecedented segmentation. Such imaging calorimeters offer additional advantages, such as the possibility to apply software compensation corrections or corrections for leakage out the back of the calorimeter. A strong R&D program is being pursued to develop such calorimeters. Several large scale prototypes have already been built by the CALICE collaboration [6] and have undergone testing in particle beams:

-    Silicon – Tungsten electromagnetic calorimeter with $1 \times 1$ cm$^2$ pads. This was the first calorimeter to take digital pictures of hadronic showers.

- Scintillator – Steel hadronic calorimeter with $3 \times 3 \rightarrow 12 \times 12$ cm$^2$ pads. This calorimeter was the first to use SiPM on a large scale.
- RPC – Steel hadronic calorimeter with $1 \times 1$ cm$^2$. This calorimeter was the first to embed its front-end electronics into the calorimeter. It also constituted the first large scale digital (= 1-bit) hadronic calorimeter. The number of readout channels approached 500,000 and represents an absolute record in calorimetry.
- RPC – Steel hadronic calorimeter with $1 \times 1$ cm$^2$. This calorimeter utilizes a semi-digital readout with 2-bit readout.

After successful testing of the first large scale prototypes, additional R&D will lead to so-called technical prototypes, which will address all technical issues related to the design and engineering of a module of a calorimeter in a colliding beam detector. Several such technical prototypes are current being built:

- Silicon – Tungsten electromagnetic calorimeter with $0.25 \times 0.25$ cm$^2$ pads.
- Silicon – Tungsten electromagnetic calorimeter with hexagonal pads with an area of 0.13 cm$^2$.
- Scintillator – Steel hadronic calorimeter with $3 \times 3$ cm$^2$ pads.

## Ackowledgments

The author would like to thank the organizers for the invitation to present recent developments in imaging calorimetry at this conference and for an impeccably organized meeting.